\documentclass[9pt,twocolumn,twoside]{article}
\usepackage{notosa}

\title{Shack-Hartmann sensor as an imaging system with a phase diversity}

\author[1,2,3,*]{Oleg Soloviev}
\author[1,4]{Hieu Thao Nguyen}
\author[3]{Vitalii Bezzubik}
	\author[3]{Nickolai Belashenkov}
\author[1,2,3]{Gleb Vdovin}
\author[1]{Michel Verhaegen}

\affil[1]{DCSC, TU Delft, Mekelweg 2, 2628 CD Delft, the Netherlands}
\affil[2]{Flexible Optical B.V., Polakweg 10-11, 2288 GG Rijswijk, the Netherlands}
\affil[3]{ITMO University, Kronverksky 49, 197101 St Petersburg, Russia}
\affil[4]{Department of Mathematics, School of Education, Can Tho University, Can Tho, Vietnam}

\affil[*]{Corresponding author: oleg.soloviev@gmail.com}

\begin{abstract}
Conventional methods of wavefront reconstruction from the raw data of the Shack-Hartmann sensor use the focal spot shifts and discard the high-frequency information about the wavefront. 
Phase-retrieval-based methods treat the Hartmann pattern as the diffraction  image and use the Rayleigh-Sommerfeld propagation to estimate the wavefront with greater accuracy and resolution.
In this Letter, we propose a novel approach to the phase-retrieval-based reconstruction by considering the Hartmann pattern as a point-spread function of a general imaging system with an introduced phase diversity of a special type. 
This model allows one not only to use \emph{any} phase retrieval algorithm to reconstruct the wavefront, but also to analyse the limitations of the phase-retrieval-based methods. 
We demonstrate the validity of this approach both on the simulated and experimental data.
\end{abstract}

\setboolean{displaycopyright}{true}

\usepackage{OlegDefinitions}
\usepackage{todonotes}

\begin{document}

\maketitle

%%%%%%%%%%%%%%%%%%%%%%%%%%  body  %%%%%%%%%%%%%%%%%%%%%%%%%%

%\section{Introduction}

Shack-Hartmann sensor~\cite{Platt2001HistoryAP}  is a widely used wavefront sensor. 
As noted by C.Dainty, despite (or because) of its intuitively clear operational principle, its potential is not understood completely. 
This might explain the vivid research activity in this field---search of the keywords ``Shack-Hartmann'' in titles and abstracts of OSA Publishing database returns more than 30 results per year in average for the last decade.

A significant part of the literature investigates the ways to increase the accuracy, dynamical range, and  resolution of the Shack-Hartmann wavefront sensor (SHWS) by modifying its construction or restoration algorithms. 
While this might be of less importance for the SHWS use in a closed loop of an adaptive optical system, there are applications like optical tests~\cite{Urbach2012,Li2015} or optical-path-difference microscopy~\cite{Gong2017} that would benefit from such improvement.

Large group of methods uses the conventional way of the wavefront reconstruction from a (Shack\mbox{-)}Hartmann pattern (HP)  which  subdivides it to the areas containing individual focal spots, then finds the centroids of the spots, and relates the shifts of the centroids to the averaged over subapertures gradient of the wavefront. 
Then the accuracy can be increased only by more precise estimation of the focal spot shifts~\cite{Thomas2006, Wei2013}, and
this approach limits SHWS to the low-order modes---the resolution of the reconstructed phase is given by the dimensions of the microlens array (MLA).
Even if the HP is processed globally (as in Ribak’s method~\cite{Talmi2004, Canovas2007}), only the information about averaged slopes is extracted, and the SHWS resolution is not increased.

Recently, several articles appeared where attempts had been made to extract the additional information from each of the HP spots, \emph{e.g.} using their second moments to retrieve the local  wavefront curvature\cite{Viegers2017},  or to run localized phase fitting algorithms
\cite{Brunner:17} and stitch the individual patches
to obtain the high-order frequencies.

Finally, the phase-retrieval-based approach keeps all available in HP information by considering it as a diffraction pattern produced by MLA.
In general, these methods use the Rayleigh-Sommerfeld transfer function to relate the complex amplitudes in the pupil and focal planes of the MLA and to reduce the wavefront reconstruction to the phase retrieval (PR) problem.
Refs.~\cite{Urbach2012, Brunner2014} use modal decomposition of the wavefront and find the best fit solution of the PR problem.
Ref.\cite{Li2014} uses modal reconstruction too and proposes to modify the sensor by moving the imaging plane out of focus which improves the PR results. 
Ref.~\cite{Li2015} registers intensities with and without the MLA and propagate the field  back and forth to provide more accurate measurements of near-flat wavefronts.
Refs.\cite{Yazdani2017} have used several detection planes and demonstrated the possibility of reconstructing vortexes in the wavefront.
All papers above demonstrated significant increase in accuracy of the estimated wavefront and analysed the convergence of the algorithms and their noise robustness, but did not investigate the dependence of the improvement on the physical parameters of SHWS, like camera pixel size and MLA pitch and focal length.

%\olegremark{To show that the approach can be used also for aberration with higher amplitude and to extend  li2015, we propose the sh diversity. and to make it possilbe for Hai's microscope}
The goal of this Letter is to provide an alternative model for the PR-based approach which makes it more clear when the phase high order information indeed can be extracted from the HP.
%
%\olegremark{Stress that the difference is just in the imaging model -- normal lens with introduced phase diversity}
%
To this end, we propose to consider the HP as an (aberrated) point spread function (PSF) of an optical system with an introduced phase diversity of  special kind, represented by a piece-wise linear defocus. 
%We demonstrate the validity of this approach and present the results obtained with Gerchberg-Saxton algorithm.
%
%
%\section{Shack-Hartmann Phase Diversity}

Consider a lens that introduces a phase delay that brings a collimated beam to a focus and establish the relationship between the pupil function and the PSF:
\begin{equation}\label{eq:PSF}
I = \abs{\F \big(A \ee ^{\ii \phi}\big) }^2,
\end{equation}
where $A$ and $\phi$ are the amplitude and the phase of the field in the pupil plane, and $\F$ denotes the (2-dimensional) Fourier transform.

\begin{figure}[th!]
	\centering
	\includegraphics[width=0.99\linewidth]{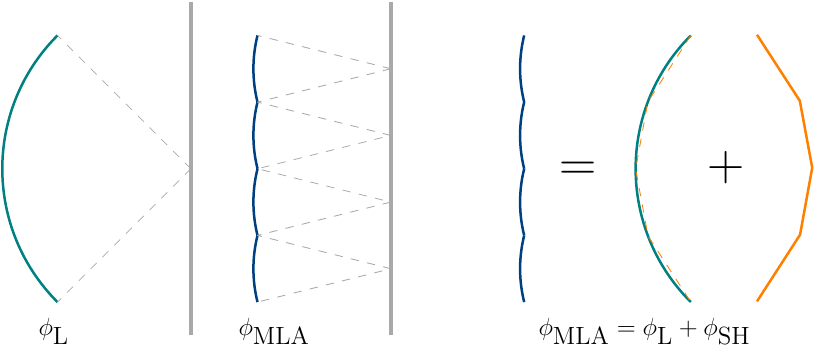}	
	\caption{The phase delays introduced by a lens ($\phi_\textrm{L}$) and by an MLA ($\phi_\textrm{MLA}$) of the same focal length are related to each other via the Shack-Hartmann diversity $\phi_\textrm{SH}$.
		The SH diversity is the linearised defocus phase with the opposite sign.}
	\label{fig:sh-diversity1d}
\end{figure}

An MLA with the same focal length introduces a phase delay consisting piece-wise of smaller copies of the same defocus.
This phase delay can be decomposed in a sum of  one global defocus  phase and piece-wise linear term  $ \phi_\textrm{SH}$ (see Fig.~\ref{fig:sh-diversity1d}), which we propose to name the Shack-Hartmann (SH) diversity.
The intensity produced with the MLA can thus be considered as  PSF of a conventional imaging system with additional SH phase diversity:
\begin{equation}\label{eq:psf-diversity}
I = \abs{\F \big(A \ee ^{\ii \phi +\ii \phi_\textrm{SH}}\big) }^2.
\end{equation} 

Unlike the formulae of Refs.~\cite{Li2014,Li2015,Brunner2014,Urbach2012,Yazdani2017}, due to the SH diversity term, \eqref{eq:psf-diversity} represents a relationship between the pupil and focal planes of a general imaging system; consequently it does not  involve  the Rayleigh–Sommerfeld equation, and requires just one Fourier transform, making conversion to the phase retrieval problem trivial. 

The idea of the SH diversity can be easily illustrated on example of the forward problem of simulation of a realistic HP
given a phase in the pupil and the MLA parameters.
A proper approach would be to isolate a portion of the phase corresponding to a given sub-aperture (s/a), and to calculate the corresponding PSF (for sake of simplicity, let us consider only low-NA case).
Independently of the s/a position, the calculated by Fourier method PSF will be located in the center of the grid, so it should be translated back under the s/a centre.
Having this done for all sub-apertures, one should add all of the translated PSF coherently and take the squared absolute value of the intensity to obtain  the final Shack-Hartmann pattern.

\begin{figure}[tbh!]
	\centering
	\includegraphics[width=0.4\linewidth]{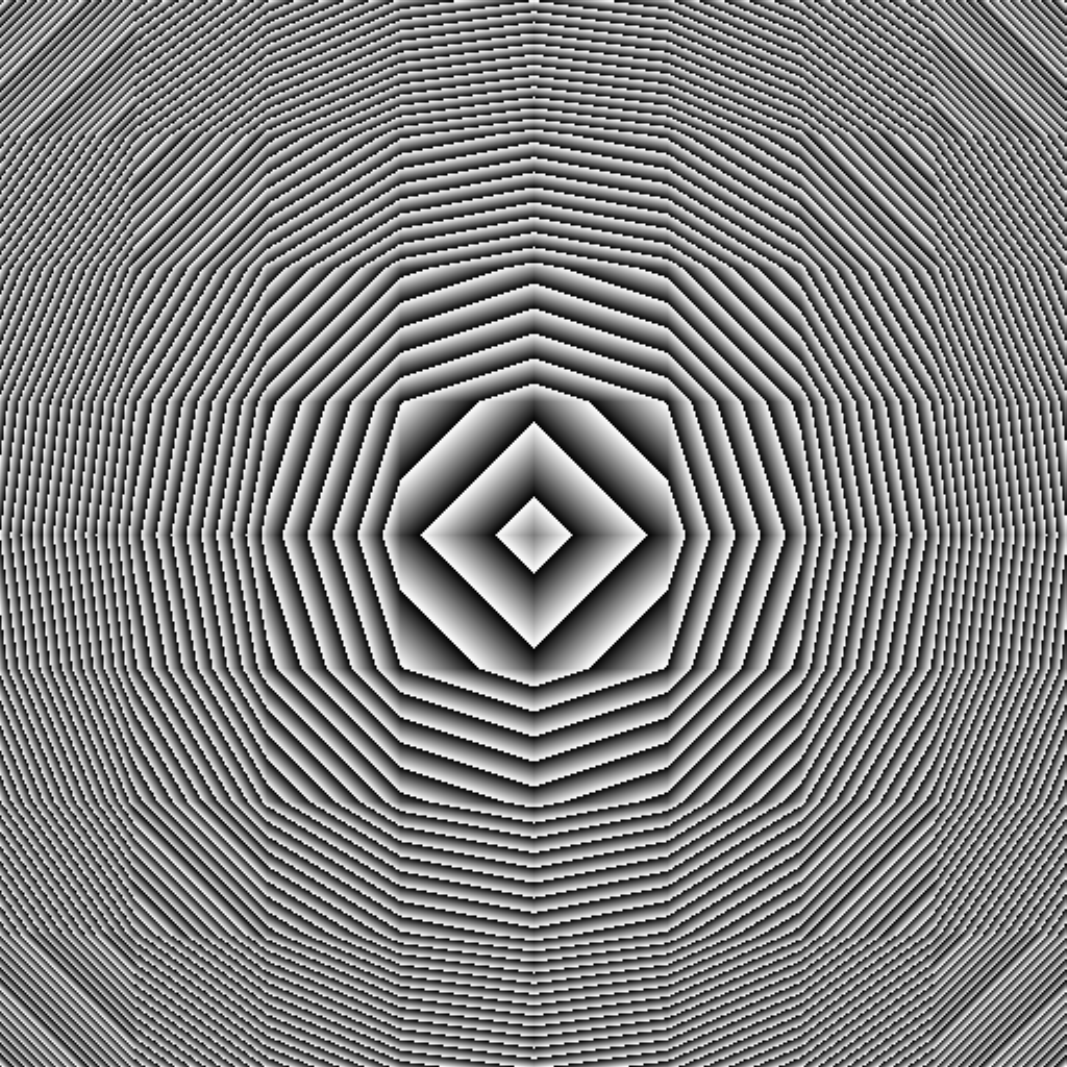}
	\includegraphics[width=0.45\linewidth]{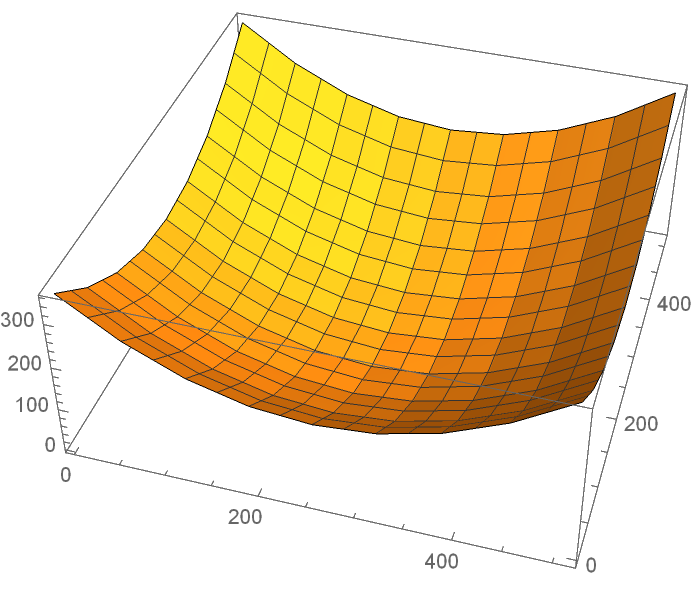}	
	\caption{Shack-Hartmann phase diversity for an $8\times 8$ microlens-array with pitch $0.15$ mm and focal length $10$ mm, shown wrapped,  and as  a continuous 3D surface sampled on a grid of $512\times 512$ points. Note the large amplitude of the SH phase and close to the Nyquist sampling rate at the edge sub-apertures. }
	\label{fig:sh-diversity}
\end{figure}

Note that the PSF-shift operation above  is equivalent to introducing corresponding linear-phase term in the s/a field.
With this operation combined for all s/a, we can consider the HP formation as PSF of the aperture field with an additional phase diversity term composed by s/a-wise linear phase as stated in \eqref{eq:psf-diversity}.
%\begin{equation}\label{eq:psf-diversity}
%	I = \abs{\F \big(A \ee ^{\ii \phi +\ii \phi_{SH}}\big) }^2,
%\end{equation} 
From this it is clear that $\phi_\textrm{SH}$ is a piece-wise, or---more precisely---sub-aperture-wise, linear function with the slope proportional to the index $(i,j) $ of subaperture $S_{i,j}$: 
\begin{equation}\label{eq: SH phase gradient}
\nabla \phi_\textrm{SH}(\vx) = c \cdot (i,j) \quad \text {for } \vx \in S_{i,j}, 
\end{equation}
for some proportionality constant $c$ depending on the parameters of the MLA and pixel size (sampling grid), see
Figure~\ref{fig:sh-diversity}.

\begin{figure}[tb!]
	\centering
	\includegraphics[width=0.19\linewidth]{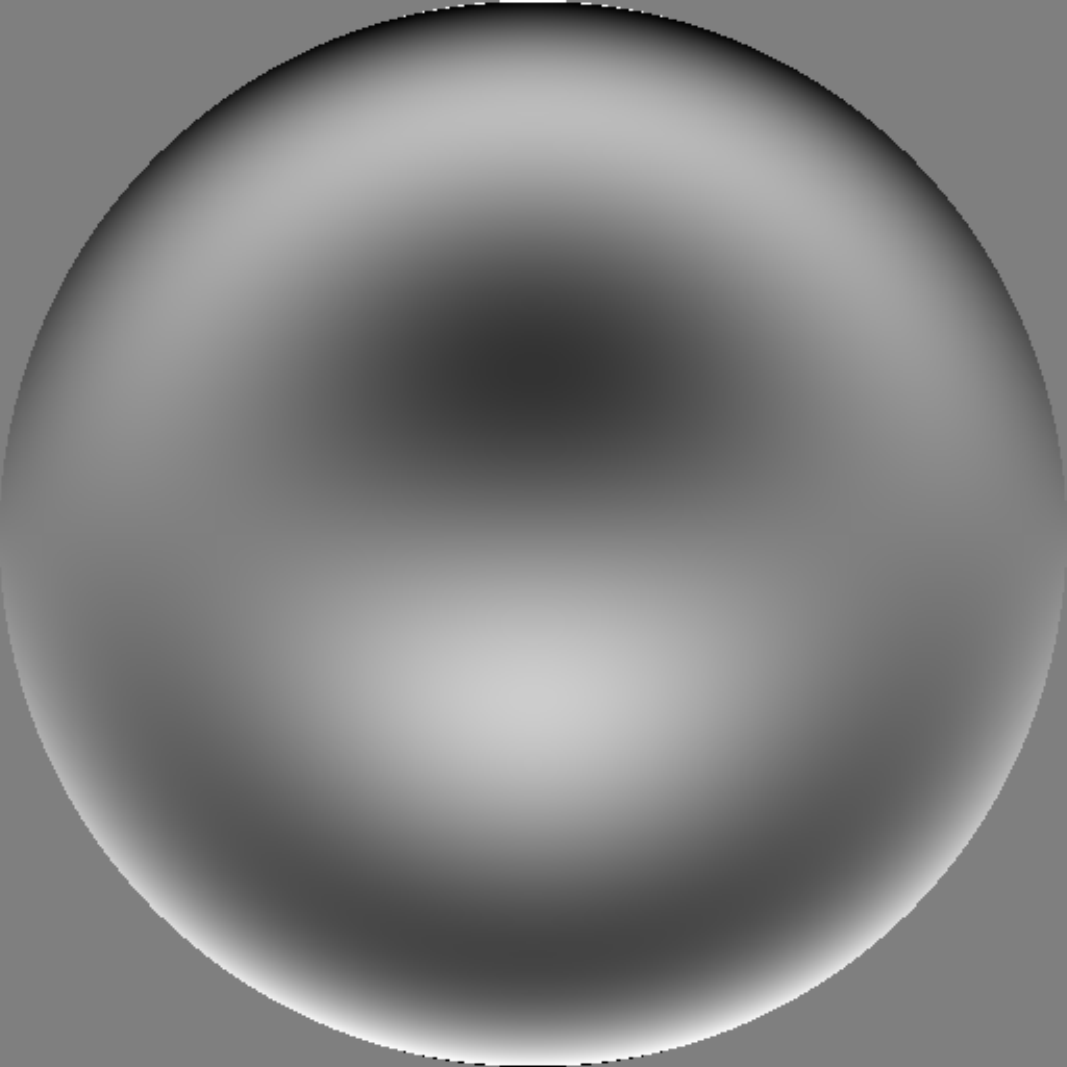}
	\includegraphics[width=0.19\linewidth]{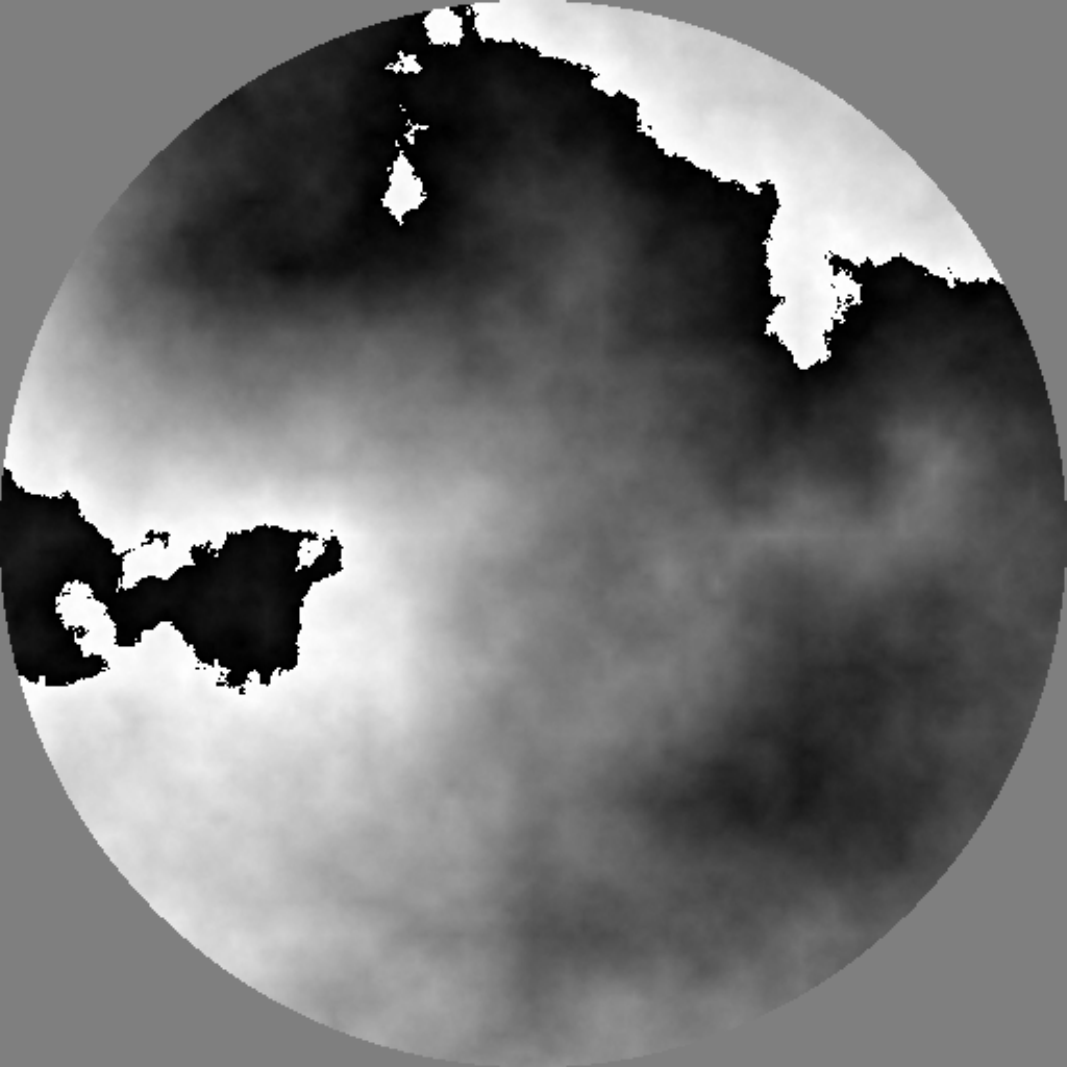}
	\includegraphics[width=0.19\linewidth]{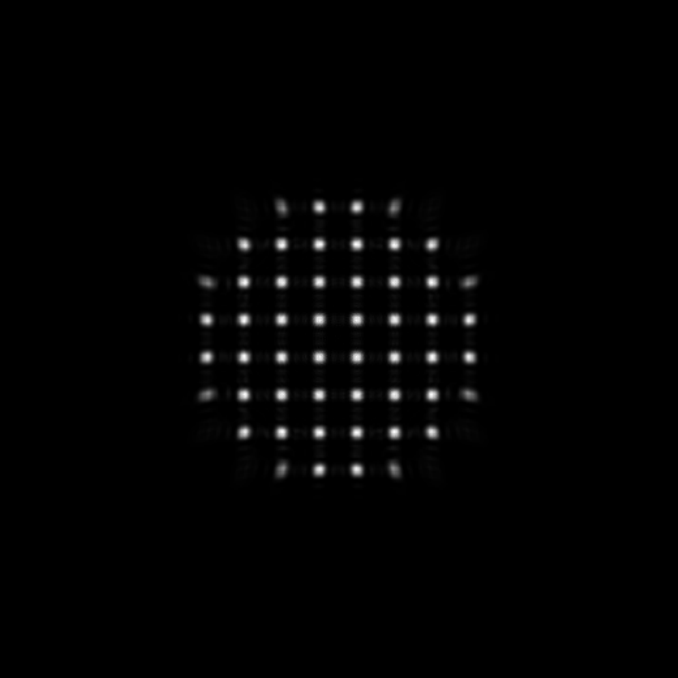}
	\includegraphics[width=0.19\linewidth]{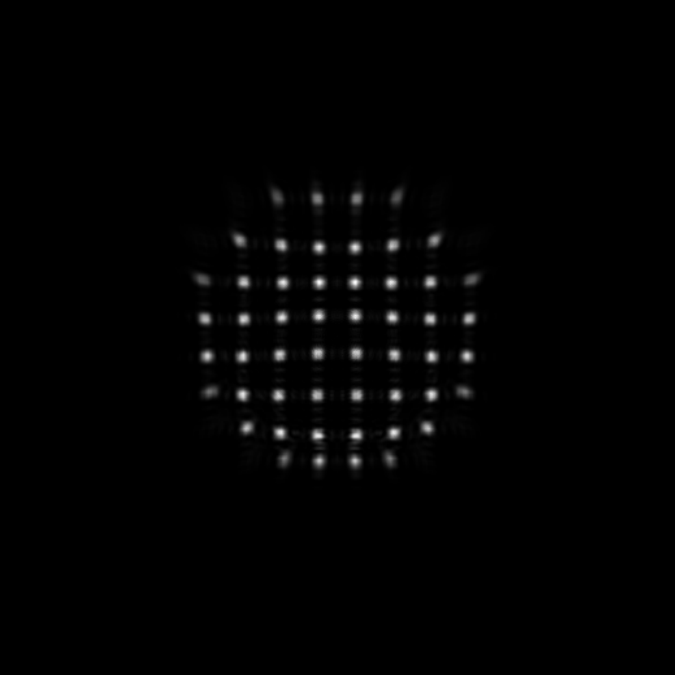}
	\includegraphics[width=0.19\linewidth]{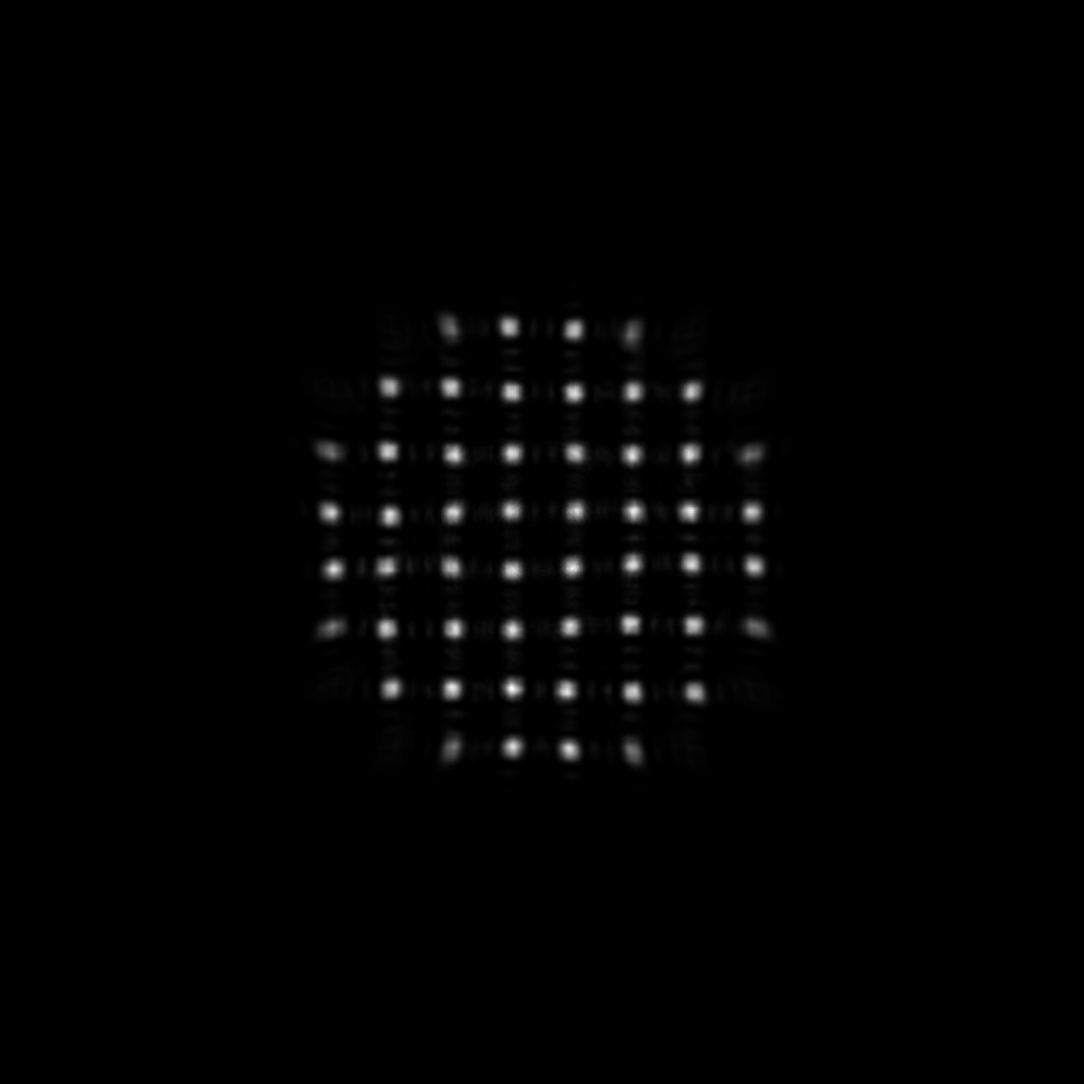}
	\caption{Phases and HP obtained from them  with method of \eqref{eq:psf-diversity} and phase diversity $\phi_{SH}$ of Fig.~\ref{fig:sh-diversity}: coma,turbulent phase, HP plane wavefront, HP coma, HP turbulent phase (left to right). The phases are shown wrapped.}
	\label{fig:sh-aberrations}
\end{figure}

This method allows for fast and accurate simulations of SH patterns, see Figure~\ref{fig:sh-aberrations} for illustration.

%\olegremark{add red reference points and the phases itself}

\begin{figure}[t!]
	\centering
	\includegraphics[width=1\linewidth]{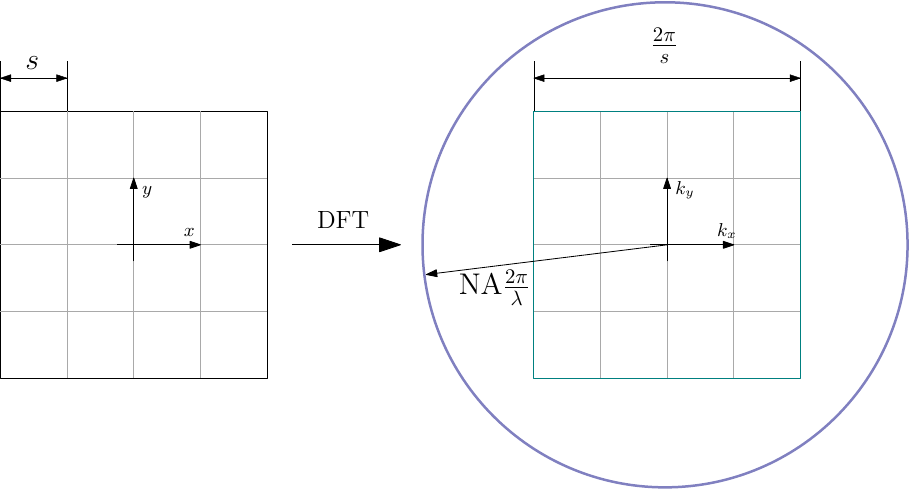}
	\caption{DFT establishes the relation between the sampling rate and the maximum frequency. In this example, pixel size $s$ is not small enough to represent all angular frequencies defined by the numerical aperture of the whole MLA; high frequencies will be aliased and not reconstructed by PR algorithms.}
	\label{fig:sampling}
\end{figure}

When simulating HP is registered with a physical device, its  parameters such as sampling rate $s$ of the camera and the light wavelength $\lambda$ should be taken into account.
By the Nyquist-Shannon principle, the maximum wavelength that can be sampled without aliasing by the camera is $2s$, which corresponds to the maximum values of the angular spectrum $k_x$ and $k_y$:
\begin{equation}\label{eq: max sample}
\abs{k_x} \leq\frac{ \pi}{s},\  \abs{k_y} \leq\frac{ \pi}{s},
\end{equation}  
while the clear aperture of the MLA introduces the limit 
\begin{equation}\label{eq: max NA}
\abs{\vecc{k}} \leq \text{NA} \frac{2 \pi}{\lambda}.
\end{equation}
For an adequate simulation of HP, the pixel size should be small enough so the following holds (see Fig.~\ref{fig:sampling}):
\begin{equation}\label{key}
\text{NA} \frac{2 \pi}{\lambda} \leq \frac{ \pi}{s}.
\end{equation}  

\begin{figure}[t!]
	\centering
	
	\includegraphics[height=0.49\linewidth]{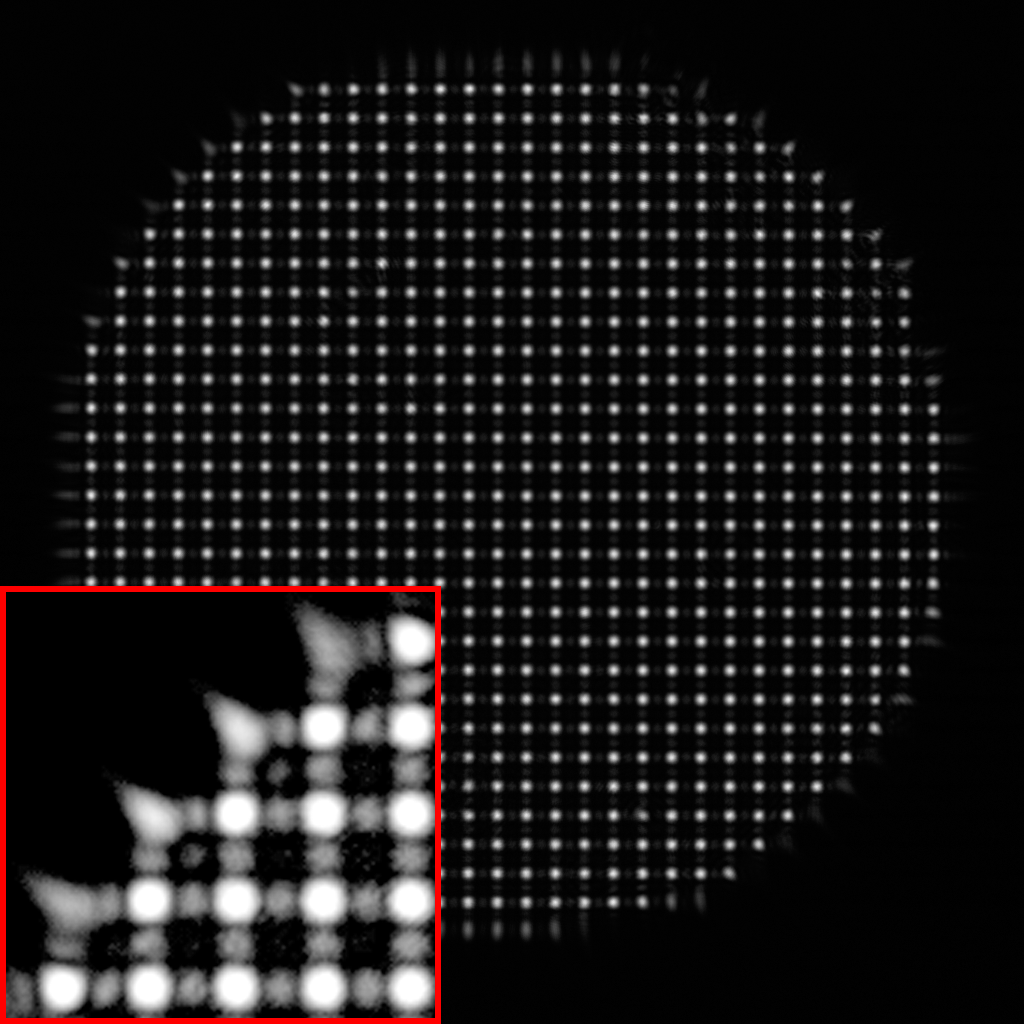}
	\includegraphics[height=0.49\linewidth]{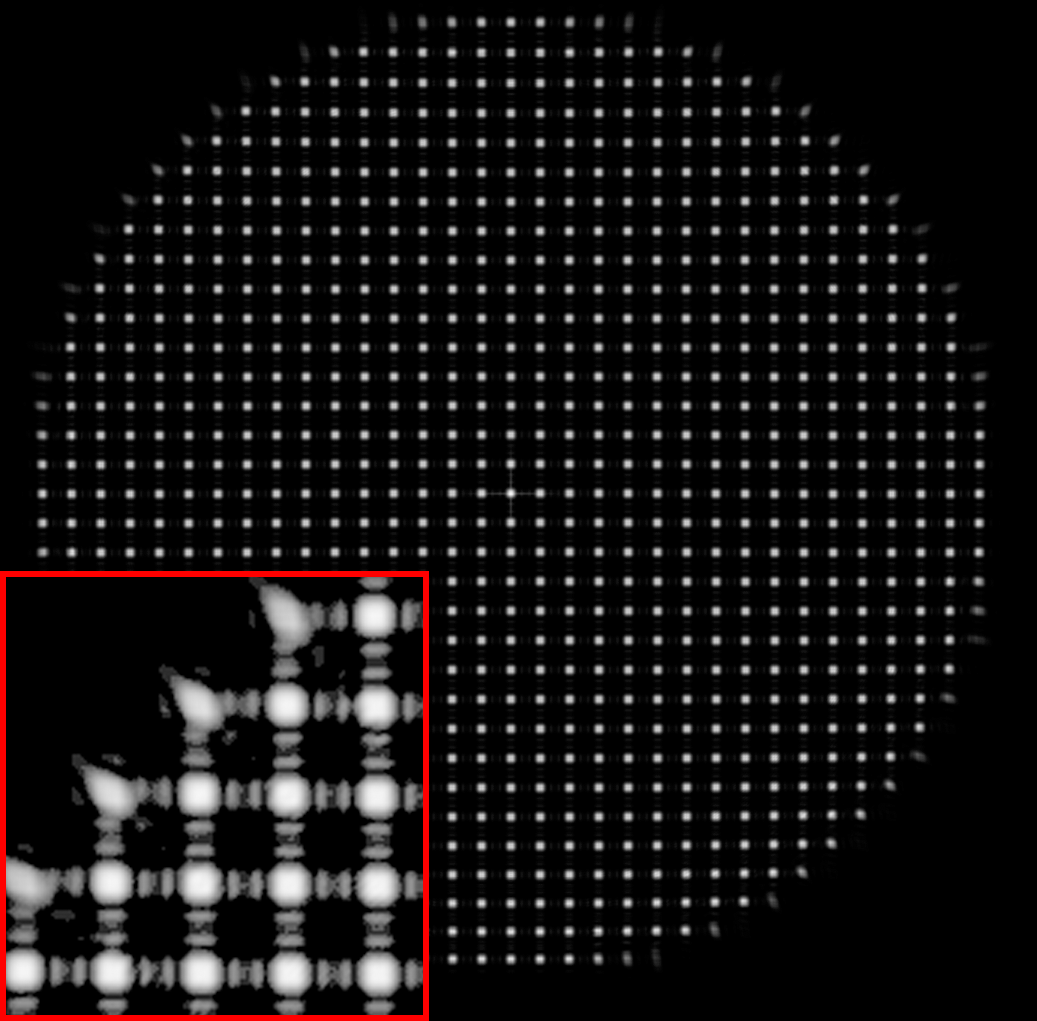}
	\caption{HP produced by an MLA with 150um pitch, 10mm focal length and $\approx$4.5 mm circular aperture: experimental (left) and simulated (right). In insets, top left fragments of HP are shown in log scale to demonstrate the under-sampling in the experimental HP.  }
	\label{fig:fullhscoma1}
\end{figure}

Consider the example of Fig.~\ref{fig:fullhscoma1}, which uses the same parameters as the  array of  Fig.~\ref{fig:sh-diversity}, but with the total aperture of $\approx$4.5 mm.
For $\lambda=633$ nm, the sampling rate of the camera ($s = 5.2\mu$m) would allow an adequate representation of the angular frequencies of the NA$\leq \frac{\lambda}{2s} \approx 0.06$, which corresponds to the maximum aperture of 1.2mm.
It follows that the HP should be simulated with at least 4 times smaller pixel size, and although the whole HP can fit in an array of about 875 pixels wide, the simulation grid for the pupil field and the SH diversity should be about 3500$\times$ 3500, and the resulted array representing the PSF should be decreased to the original size afterwards by decimation or by averaging.
In addition, the under-sampling introduces  aliasing of the higher frequency components in the experimental HP, and they cannot be recovered without additional \emph{a priori} knowledge.

%\section{PHASE RETRIEVAL FROM SH PATTERNS}
Equation (\ref{eq:psf-diversity}) represents a SH pattern in a form of a PSF, which can be converted to the phase retrieval problem
\begin{equation}\label{eq: PR}
\text{find } X\in \Complex^{N\times N} \ \text{s.t.  } \	X = \abs{\F x},
\end{equation} 
and in case of a proper sampling we can directly apply any of the existing phase-retrieval methods (see~\cite{Shechtman2015,Luke2017} for an overview).
For simplicity, we have chosen the Gerchberg-Saxton (GS) algorithm.

\begin{figure}[t!]
	\centering
	\includegraphics[width=0.24\linewidth]{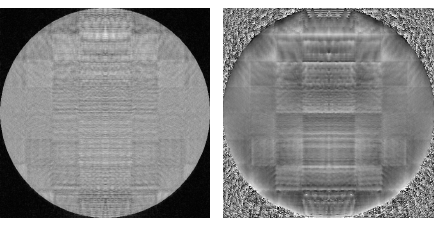}\hfill
	\includegraphics[width=0.24\linewidth]{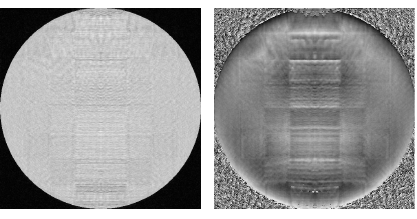}\hfill
	\includegraphics[width=0.24\linewidth]{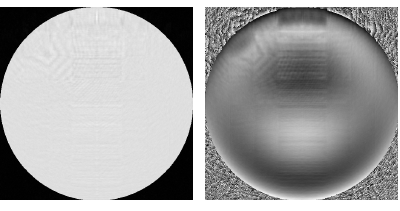}\hfill
	\includegraphics[width=0.24\linewidth]{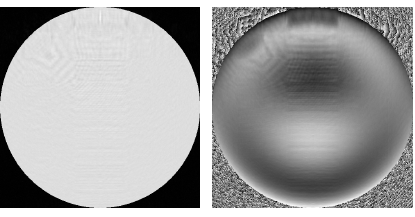}\\
	\includegraphics[width=0.24\linewidth]{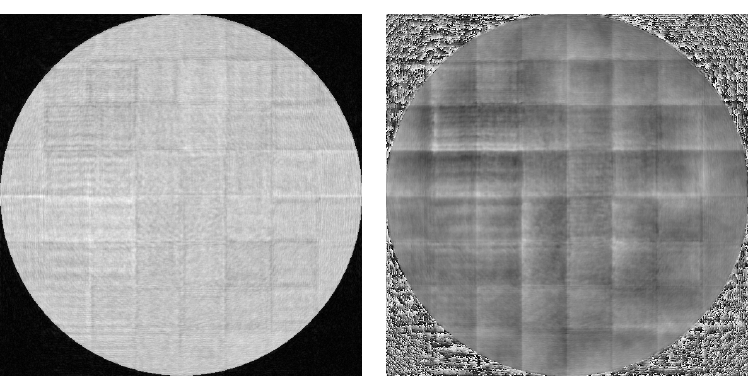}\hfill
	\includegraphics[width=0.24\linewidth]{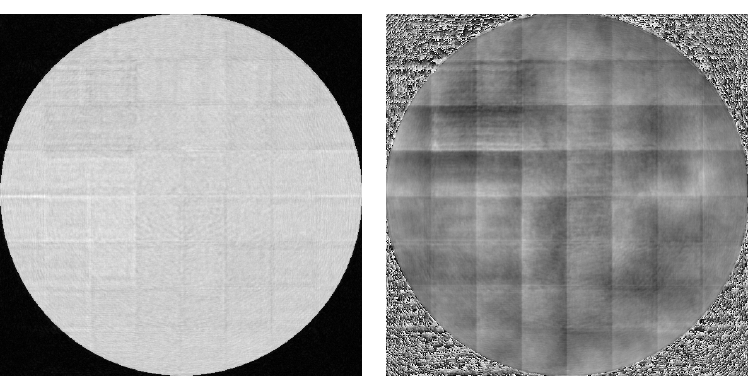}\hfill
	\includegraphics[width=0.24\linewidth]{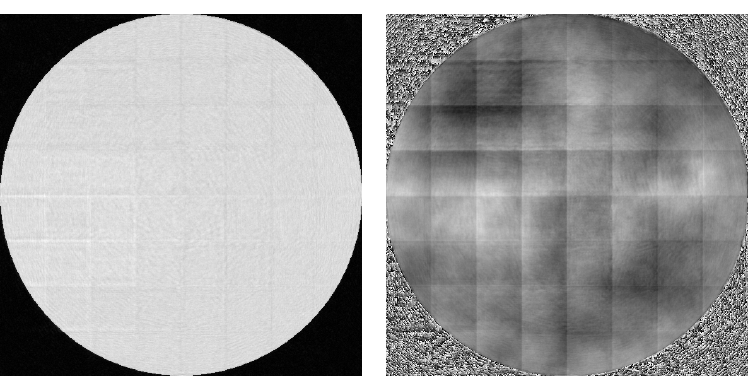}\hfill
	\includegraphics[width=0.24\linewidth]{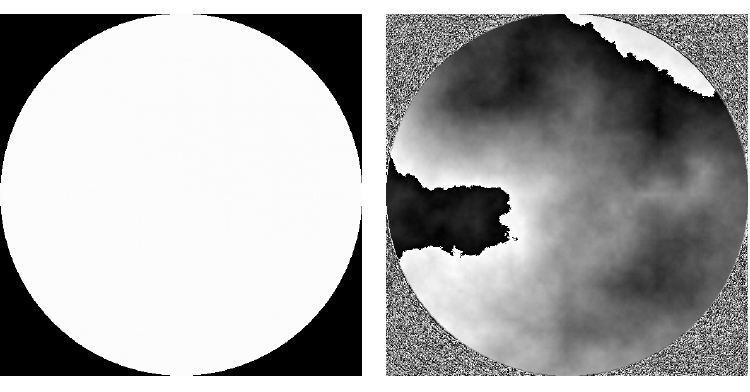}
	\caption{Amplitude and phase restored by the GS algorithm from the SH patterns of Figure~\ref{fig:sh-aberrations} after 10, 100, 1000, and 10000 iterations for coma (top) and turbulent (bottom) phases. The artefacts in the upper part of the restored coma phase are caused by (almost) under-sampling.}
	\label{fig:iterations}
\end{figure}

\begin{figure}[bth!]
	\centering
	\includegraphics[ height=.3\linewidth]{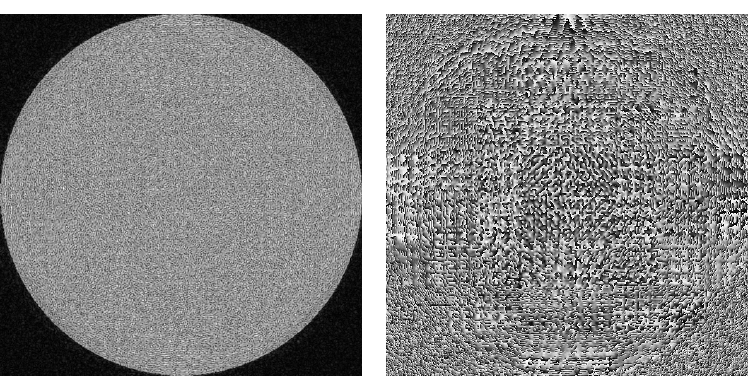}\hspace{5mm}
	\includegraphics[ height=.3\linewidth]{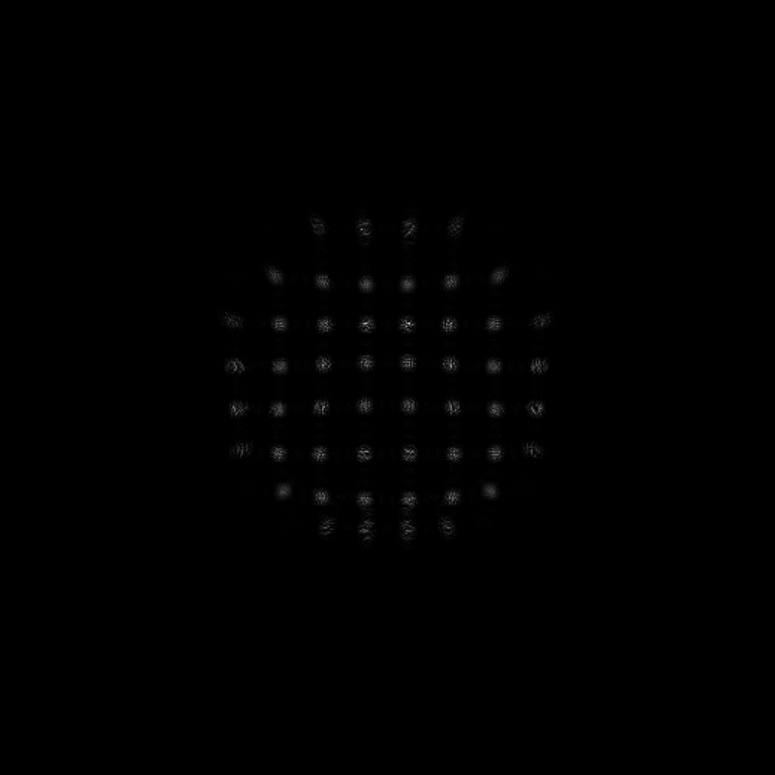}
	\caption{Wrong solution provided by the GS method initiated by a zero phase, from left to right: amplitude and phase of the pupil field, SH pattern.}
	\label{fig:comafrom0}
\end{figure}

The solution provided by a GS method is known to be dependent on the initial point. 
Considering the relative magnitude of the phase terms in \eqref{eq:psf-diversity}, it seems reasonable to set $\phi_\textrm{SH}$ as the initial point of the algorithm. 
(It is interesting to note that, on the contrary, starting from a zero phase results in a wrong solution with speckle-like HP, as shown in Fig.~\ref{fig:comafrom0}).
The results obtained after various number of iterations of the GS phase retrieval from HP obtained from a smooth and turbulence-like phases  are shown in Fig.~\ref{fig:iterations};
from them it is obvious that the phase can be restored from the noiseless SH patterns at every grid position if the Nyquist sampling criterium was satisfied and total number of iterations is big enough: for instance, the \emph{rms} error of the restored turbulent phase after 20 000 iterations was below $\lambda/10^5$.

\begin{figure}[b!]
	\centering
	\includegraphics[width=1\linewidth]{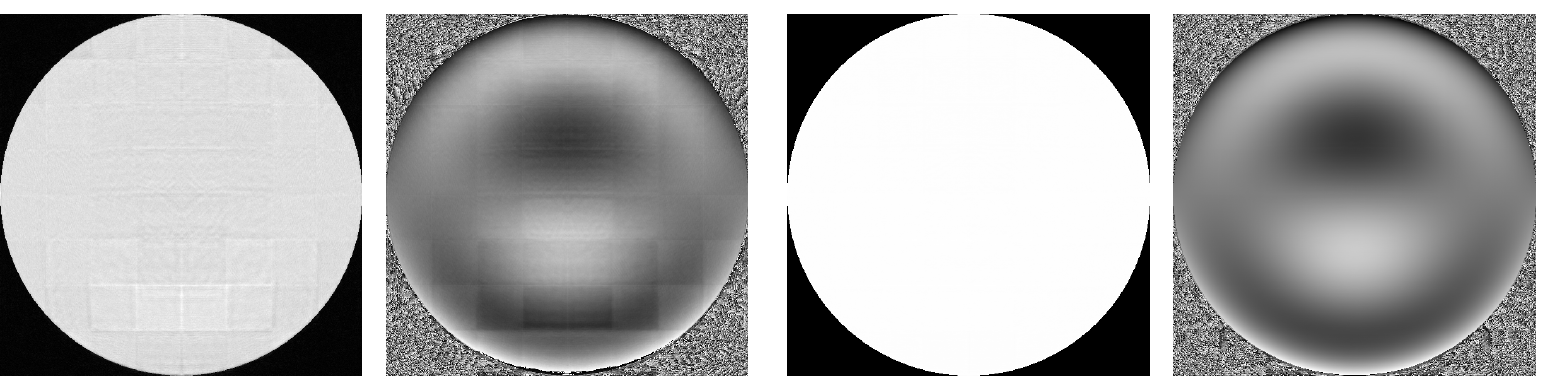}
	\caption{Amplitude and phase restored by the GS method using $\phi_\textrm{WFS} + \phi_\textrm{SH}$ as starting point after 10 and 1000 iterations.}
	\label{fig:coma10-1000-fs}
\end{figure}

The total number of iterations can be reduced and the accuracy of the solution can be increased if we use additional information provided by the SH pattern---namely, the phase $\phi_\textrm{WFS}$ obtained by the traditional slope-based method.
Figure~\ref{fig:coma10-1000-fs} shows the results of the GS method with $\phi_\textrm{WFS}+\phi_\textrm{SH}$ as initial phase.

%While it is known that the solution of the phase retrieval problem by alternative projections methods does strongly depends on the initial point due to non-convexity of the problem, in the proposed method it is crucial to begin from a phase close to the SH diversity in order to converge to a proper solution.
%A good \emph{a posteriori} criterion on the correctness of the restored phase is the closeness of the restored amplitude to the (known) value---see, for instance, the development of the amplitude in Fig.~\ref{fig:coma10-1000-fs}.
%This criterion allows to classify the solution of Fig.~\ref{fig:comafrom0} as wrong one.

\begin{figure}[tbh!]
	\centering
	\includegraphics[width=0.5\linewidth]{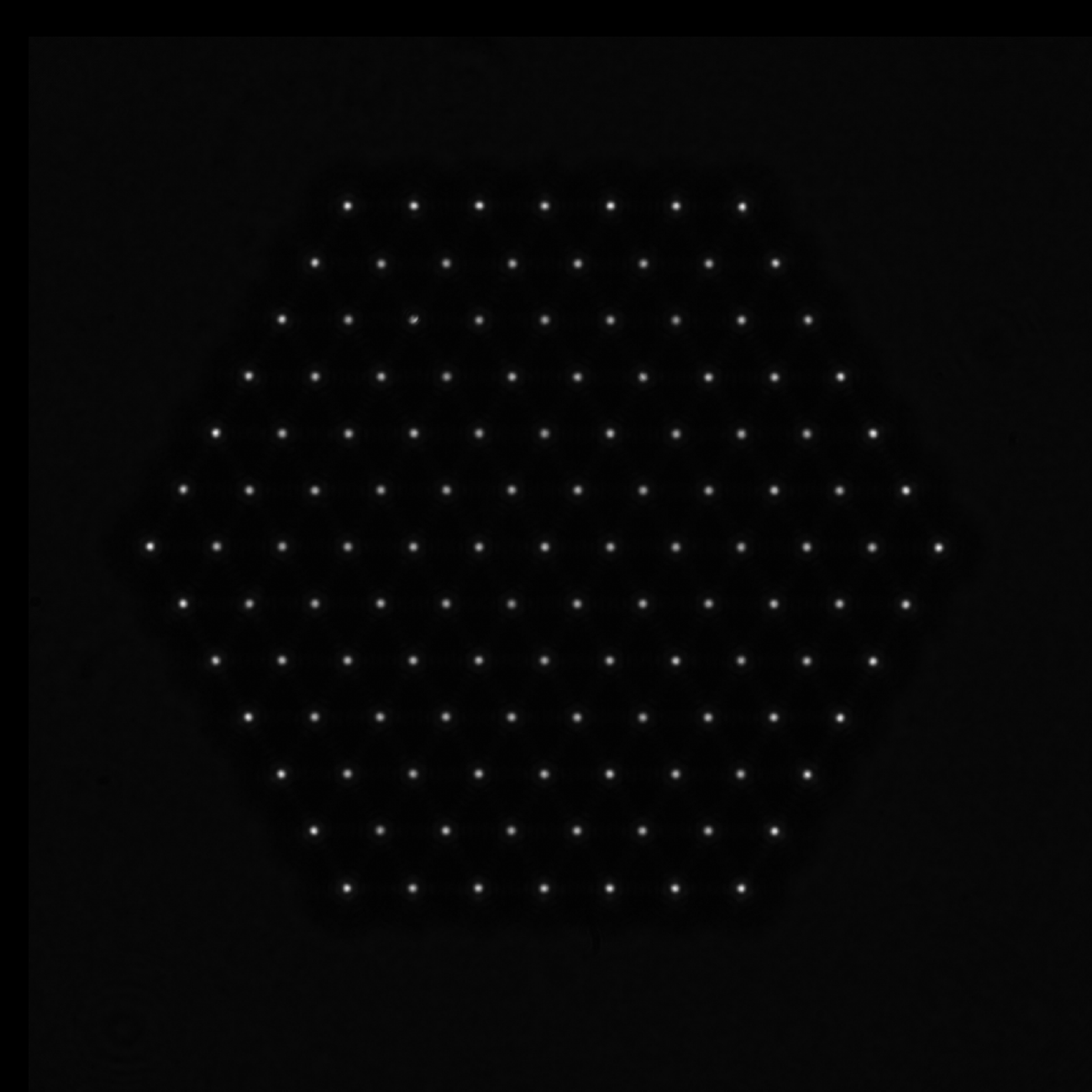}
	\caption{Experimental HP obtained with a hexagonal MLA consisting of 127 microlenses with pitch of 300$\mu$m and focal length of 18mm, registered with UI-1540 camera (pixel size 5.2$\mu$m). The exact distance of the MLA from the camera chip and its rotation were considered as unknown during the processing.}
	\label{fig:hexsh}
\end{figure}

%\section{Real Shack-Hartmann patterns}
To demonstrate some possible implications related to the (under\mbox{-)}sampling of the HP and need for the correct initial point in the absence of the exact geometrical pattern as in Fig.~\ref{fig:sh-diversity}, we applied the proposed method to the HP obtained with a hexagonal MLA (OKO Tech) with 127 microlensess with pitch of 300$\mu$m, and focal length of 18mm, as shown in Fig.~\ref{fig:hexsh}.
%According to the manufacturer, the parameters of the MLA and the camera are as specified in the table...
Note that while exact distance and rotation angle of the MLA with respect to the camera chip are specified by the manufacturer, in the following tests they were considered as unknown to demonstrate a nice auto-calibration feature of the proposed method.

%Here we can see from the data of the table, that the HP is heavily undersamped, as from the pixels size that we have and the wavelength, the maximum aperture size is about 1.2 mm (corresponding to the aperture sizes of the previous section). 
%As our aperture is 4 times larger, we need at least 4 times smaller pixel.

Although the $f$-number of the microlenses is similar to the  simulation example above, due to a longer focal distance we need only 2 times smaller pixel size for the adequate representation of the angular spectrum, but unlike the simulation, the pixel data of Fig.~\ref{fig:hexsh} is the all information we have, and half of the frequencies is lost.
One of the possible approaches to deal with this under-sampling is to interpolate the data to 2 times larger resolution.  Note that any interpolation in fact already uses some \emph{a priori} assumptions.

For the GS algorithm, we need also to know the aperture of the MLA.
As it is also not known exactly, we start with the ``default'' initial point as described below and demonstrate how the use of \emph{a priori} knowledge on the shape of the SH diversity can be used to automatically recover it.

\begin{figure}[tbh!]
	\centering
	\includegraphics[width=0.99\linewidth]{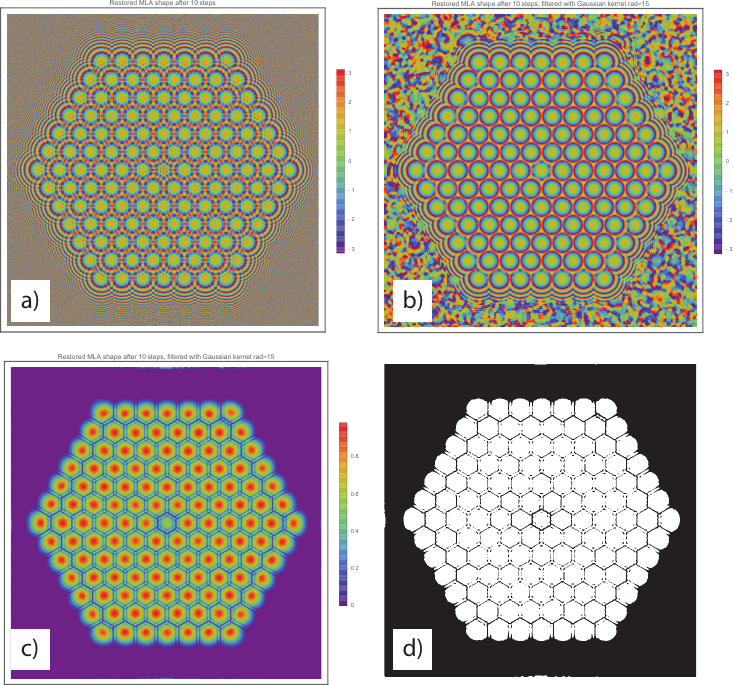}
	\caption{a) Phase delay of the MLA restored after 10 iterations; b) the same phase filtered with Gaussian kernel; c) Absolute value of the complex amplitude filtered with Gaussian kernel and d) binarised}
	\label{fig:mlaafter10steps}
\end{figure}

\begin{figure}[tbh!]
	\centering
	\includegraphics[width=0.99\linewidth]{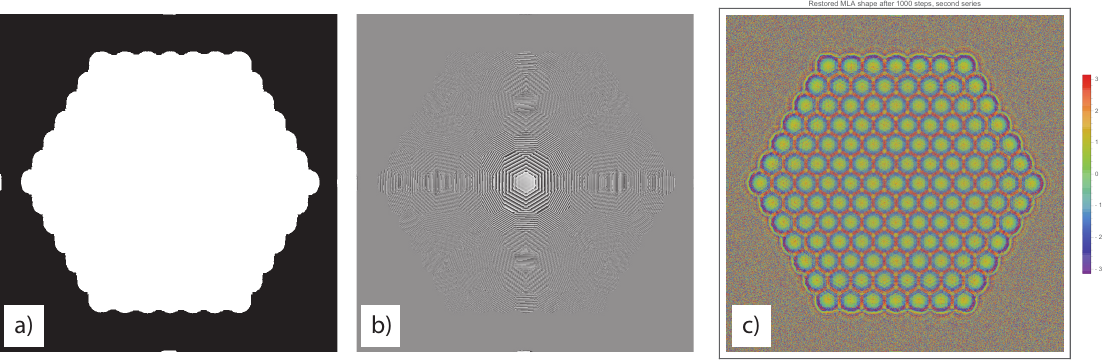}
	\caption{a) Absolute value and b) phase of the second initial point obtained using the proposed method  and c) the phase delay of the MLA reconstructed after 1000 iterations starting from the second initial point}
	\label{fig:field1}
\end{figure}

We started with the first initial point of  a circular aperture of the specified 4.2 mm and smooth (not linearised) defocus satisfying \eqref{eq: SH phase gradient}, and the GS algorithm quickly got stuck in some local minimum.
We took the solution provided by the algorithm after 10 iterations%
%($\hat\phi_\text{SH}$)
, and added the common defocus to it to recreate the MLA shape (Fig.~\ref{fig:mlaafter10steps}a).
Note how for the frequencies missing in the HP, GS restores a random phase, so
averaging with a Gaussian window the restored complex amplitude reduces to zero values in the area outside the aperture (Fig.~\ref{fig:mlaafter10steps}c).
Using sequentially binary thresholding (Fig.~\ref{fig:mlaafter10steps}d) and morphological closing operation restores the aperture shape which will be used as the modulus of the second initial point (Fig.~\ref{fig:field1}a).
For the phase of the second intitial point, we took the Gaussian average of the restored MLA phase (Fig.~\ref{fig:mlaafter10steps}b) with the subtracted common defocus. 
Starting from this second initial point, the GS algorithms converged to a proper solution (Fig.~\ref{fig:field1}c). 

In conclusion, we have proposed to model the intensity distribution of a Shack-Hartmann pattern as a PSF of a lens with an introduced piece-wise linear defocus phase (Shack-Hartmann diversity).
This allows to use the developed PSF phase retrieval methods to reconstruct the wavefront from the HP and to analyse the validity of such reconstruction based on the physical parameters of the Shack-Hartmann sensor.
In addition, we have demonstrated the auto-calibration feature of the method by restoring the Shack-Hartmann diversity from an experimental data obtained with a hexagonal microlens array.

\section*{Funding}
%Please identify all appropriate funding sources by name and contract number. Funding information should be listed in a separate block preceding any acknowledgments. List only the funding agencies and any associated grants or project numbers, as shown in the example below:\\
%\\
%National Science Foundation (NSF) (1253236, 0868895, 1222301); Program 973 (2014AA014402); Natural National Science Foundation (NSFC) (123456).\\
%\\
%OSA participates in \href{https://www.crossref.org/fundingdata/}{Crossref's Funding Data}, a service that provides a standard way to report funding sources for published scholarly research. To ensure consistency, please enter any funding agencies and contract numbers from the Funding section in Prism during submission or revisions.

The research at ITMO University is carried out  under the financial support of the Ministry of Science and Higher Education of the Russian Federation.

%\section*{Acknowledgments}
%Acknowledgments, if included, should appear at the end of the document. The section title should not be numbered.

\section*{Disclosures}
The authors declare no conflicts of interest.

%\section{Conclusion}

%%%%%%%%%%%%%%%%%%%%%%% References %%%%%%%%%%%%%%%%%%%%%%%%%

%%%%%%%%%% If using BibTeX:
%\bibliography{sample}
\bibliography{ref} 
\bibliographyfullrefs{ref}

%%%%%%%%%% If preparing manually:
% \begin{thebibliography}{1}
% \newcommand{\enquote}[1]{``#1''}

% \bibitem{Zhang:14}
% Y.~Zhang, S.~Qiao, L.~Sun, Q.~W. Shi, W.~Huang, L.~Li, and Z.~Yang,
%   \enquote{Photoinduced active terahertz metamaterials with nanostructured
%   vanadium dioxide film deposited by sol-gel method,}
%   {\protect\JournalTitle{Optics Express}} \textbf{22}, 11070--11078 (2014).

% \bibitem{OSA}
% {Optical Society}, \enquote{{OSA Publishing},}
%   \url{http://www.osapublishing.org}.

% \bibitem{FORSTER2007}
% P.~Forster, V.~Ramaswamy, P.~Artaxo, T.~Bernsten, R.~Betts, D.~Fahey,
%   J.~Haywood, J.~Lean, D.~Lowe, G.~Myhre, J.~Nganga, R.~Prinn, G.~Raga,
%   M.~Schulz, and R.~V. Dorland, \enquote{Changes in atmospheric consituents and
%   in radiative forcing,} in \enquote{Climate Change 2007: The Physical Science
%   Basis. Contribution of Working Group 1 to the Fourth assesment report of
%   Intergovernmental Panel on Climate Change,}  S.~Solomon, D.~Qin, M.~Manning,
%   Z.~Chen, M.~Marquis, K.~B. Averyt, M.~Tignor, and H.~L. Miler, eds.
%   (Cambridge University Press, 2007).

% \end{thebibliography}

\end{document}